\documentclass[twocolumn,prl]{revtex4}

\usepackage{dcolumn}
\usepackage{amsmath}

\usepackage{graphicx}
\usepackage{pslatex}

\newcommand{\e}{\mathrm{e}}

\newcommand{\av}[1]{\langle#1\rangle}
\newcommand{\eref}[1]{(\ref{#1})}

\newcommand{\defn}{\textit}
\newcommand{\Li}{\mathop{\mathrm{Li}}\nolimits}

\newlength{\figurewidth}
\begin{document}

\title{Threshold effects for two pathogens spreading on a network}
\author{M. E. J. Newman}
\affiliation{Department of Physics and Center for the Study of Complex
Systems,\\
University of Michigan, Ann Arbor, MI 48109--1120}
\affiliation{Santa Fe Institute, 1399 Hyde Park Road, Santa Fe, NM 87501}
\begin{abstract}
Diseases spread through host populations over the networks of contacts
between individuals, and a number of results about this process have been
derived in recent years by exploiting connections between epidemic
processes and bond percolation on networks.  Here we investigate the case
of two pathogens in a single population, which has been the subject of
recent interest among epidemiologists.  We demonstrate that two pathogens
competing for the same hosts can both spread through a population only for
intermediate values of the bond occupation probability that lie above the
classic epidemic threshold and below a second higher value, which we call
the coexistence threshold, corresponding to a distinct topological phase
transition in networked systems.
\end{abstract}
\pacs{}
\maketitle

Social, technological, and biological networks of various kinds have been
the subject of a large number of recent studies published in the physics
literature~\cite{Strogatz01,AB02,DM02,Newman03d}.  One of the principle
practical applications of this body of work has been in modeling the spread
of epidemic disease.  Diseases spread over the networks of physical
contacts between individuals~\cite{Mollison77,SS88,PV01a,Newman02c} and an
understanding of the structure of these networks and the dynamics of
disease upon them is crucial to the development of strategies for disease
control.  As it turns out, a large class of epidemic processes can be
mapped onto bond percolation
models~\cite{Mollison77,Grassberger82,Sander02,Newman02c}, allowing
familiar techniques from statistical physics to be applied directly to
their solution.

An issue of some interest in current epidemiological research is the
behavior of competing pathogens~\cite{Dietz79,CHL96,ALL97,GFA98}.  Two
diseases may compete for the same population of hosts because one disease
kills hosts before the other can infect them.  Or there may be
cross-immunity between the diseases such that exposure to one disease
leaves the host alive but immune to further infection by either disease.
Competing strains of influenza can show this type of behavior, for
instance~\cite{ALL97,GFA98}.  The dynamics of competition between pathogens
is in general complex, depending in particular on whether one pathogen gets
a head start on the other in the population.  In this paper we study the
case in which two pathogens pass through the population at well separated
intervals: one infects the population and causes an epidemic, leaving some
fraction of the population immune or dead, and at some later time the
second pathogen passes through the remaining population.  (Our arguments
could also be applied to two successive outbreaks of the same disease.)
The question we address is if and when the second disease is able to
spread.  If a sufficient number of hosts is removed from the population by
the first disease then spread of the second becomes impossible.  As we will
see there is a threshold value of the bond occupation probability or
``transmissibility'' for the first disease (a measure of contagiousness) at
which this happens.  This ``coexistence threshold'' coincides with a
continuous phase transition similar to the well known epidemic transition,
but the two transitions are quite distinct: the coexistence threshold is an
additional property of the network topology.

Spread of both pathogens can occur only in the intermediate regime between
the epidemic and coexistence thresholds.  Among other things, we determine
by exact analytic calculation for a broad class of networks the position of
the two thresholds.  For the much-studied case of a ``scale-free'' network,
we find that while the epidemic threshold for the first disease is always
zero, the coexistence threshold is not.  A corollary of this result is that
while a single disease on a scale-free network cannot be eradicated solely
by lowering the transmissibility, a similar intervention in the case of two
competing diseases \emph{can} eradicate one of the diseases, but not both.

Consider then an epidemic taking place on a network of contacts between
individuals.  The network is represented by a graph in which vertices are
individuals and (undirected) edges are contacts.  The epidemic begins with
a single individual and spreads along the contacts.  Not every contact
necessarily results in disease transmission however.  We assume a
generalized susceptible/infective/removed~(SIR) dynamics for the disease of
the kind described in~\cite{Newman02c} in which the disease spreads over
edges with a probability~$T$ called the \defn{transmissibility} of the
disease.  This dynamics can be mapped onto a bond percolation process on
the same graph with bond occupation probability equal to the
transmissibility~\cite{Mollison77,Grassberger82,Sander02,Newman02c}.  The
connected clusters of vertices in the percolation process then correspond
to the groups of individuals who would be infected by a disease outbreak
starting with any individual within that cluster.  Typically, for small
values of~$T$ there are only small clusters and hence only small disease
outbreaks.  But above some critical transmissibility~$T_c$ an extensive
spanning cluster or ``giant component'' appears, corresponding to an
epidemic of the disease: once such a giant component is present, the
pathogen reaching any of its members will infect them all and thereby reach
an extensive fraction of the population.  The value of~$T$ at which the
giant component first forms is called the epidemic threshold and it
corresponds precisely to the percolation threshold for percolation on the
contact network.

To be concrete, we examine in this paper the class of graphs that have
specified degree distribution but which are otherwise random, in the limit
of large graph size.  (Recall that the ``degree'' of a vertex in a network
is the number of edges connected to that vertex.)  Such graphs have been
studied in the past by many authors~\cite{Luczak92,MR95,NSW01,CL02b,BR04}
and have become a standard arena for the exploration of epidemiological
processes~\cite{PV01a,Newman02c}.  Epidemiological processes have also been
studied on other types of networks, such as networks with degree
correlations~\cite{BP02a,CBH03}, and it seems likely that the results
presented in this paper could be generalized to such cases, although we do
not do that here.

Let $p_k$ be the fraction of vertices in our network that have degree~$k$.
We can also consider $p_k$ to be the probability that a randomly chosen
vertex has degree~$k$.  The vertex at either end of a randomly chosen edge,
on the other hand, has degree~$k$ with probability proportional not
to~$p_k$ but to~$kp_k$, the reason being that there are $k$ times as many
edges connected to a vertex of degree~$k$ than to a vertex of degree~1, and
hence the probability that our edge will be one of them is also multiplied
by~$k$.  We will primarily be interested in the distribution of the number
of edges emerging from such a vertex other than the one we followed to get
there.  This \defn{excess degree} is one less than the total degree of the
vertex and therefore has a (correctly normalized) distribution
\begin{equation}
q_k = {(k+1)p_{k+1}\over\sum_k kp_k} = {(k+1)p_{k+1}\over z},
\end{equation}
where $z=\av{k}=\sum_k kp_k$ is the mean degree of the vertices in the
network.

Our first pathogen can spread across the network if its \defn{basic
reproductive number}~$R_0$ is greater than unity, i.e.,~if for every person
infected the mean number of additional people they infect is greater
than~1.

When the disease arrives at a vertex, it has the chance to spread to any of
the $k$ other neighbors of that vertex, each of which chances is realized
with probability~$T$ for an expected $Tk$ additional vertices infected.
Averaging over the distribution~$q_k$ of~$k$, we find that the basic
reproductive number is
\begin{equation}
R_0 = T \sum_{k=0}^\infty kq_k
    = {T\over\av{k}} \sum_{k=0}^\infty k(k+1)p_{k+1}
    = T {\av{k^2}-\av{k}\over\av{k}}.
\end{equation}
Thus the disease spreads if and only if $T$ is greater than the critical
value
\begin{equation}
T_c = {\av{k}\over\av{k^2}-\av{k}}.
\label{threshold}
\end{equation}

To calculate the size of the epidemic when it does occur, it is convenient,
following our previous approach~\cite{NSW01,CNSW00,Newman02c}, to define
two probability generating functions for the distributions $p_k$ and~$q_k$:
\begin{equation}
F_0(x) = \sum_{k=0}^\infty p_k x^k,\quad
F_1(x) = \sum_{k=0}^\infty q_k x^k = {F_0'(x)\over z},
\label{defsf1}
\end{equation}
where $F_0'$ denotes the first derivative of $F_0$ with respect to its
argument.  In terms of these functions, for example, Eq.~\eref{threshold}
can be written
\begin{equation}
T_c = {1\over F_1'(1)}.
\label{tc}
\end{equation}

Now let $u$ be the mean probability that a vertex is \emph{not} infected by
a specified neighboring vertex in the network during an epidemic outbreak
of our disease.  This quantity is equal to the probability that no
transmission occurred between the two vertices, which is $1-T$, plus the
probability that there was contact sufficient for transmission but that the
neighboring vertex itself wasn't infected.  The probability that the
neighboring vertex wasn't infected is equal to the probability that it, in
turn, failed to contract the infection from any of its $k$ other neighbors,
which is just~$u^k$ with $k$ distributed according to~$q_k$.  Thus the mean
probability of the neighbor being uninfected is $\sum_{k=0}^\infty q_k u^k
= F_1(u)$.  Hence, $u$~must satisfy the equation
\begin{equation}
u = 1 - T + T F_1(u).
\label{defsu}
\end{equation}
Then the probability that a randomly chosen vertex is not infected is
$\sum_{k=0}^\infty p_k u^k = F_0(u)$, and the fraction~$S$ of vertices that
do get infected is one minus this:
\begin{equation}
S = 1 - F_0(u).
\label{gc}
\end{equation}
Thus we can calculate the size of the epidemic by solving~\eref{defsu}
for~$u$ and substituting the result into~\eref{gc}.  $S$~is also the
probability of an epidemic occurring if the disease starts with a randomly
chosen individual---with probability $1-S$ an outbreak fails to become an
epidemic even when we are above the transition.  If $S$ is regarded as an
order parameter for the model, then the epidemic transition is a continuous
phase transition in the mean-field universality class for percolation.

\begin{figure}
\includegraphics[width=7cm]{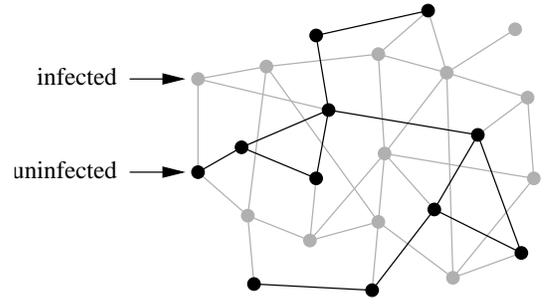}
\caption{An epidemic of the first disease spreads through the network
leaving some fraction of the vertices either immune to further infection or
dead, which we represent by simply removing them and their adjacent edges
from the network (gray).  The question we address is whether the remaining
``residual network'' (black) percolates and can therefore support the
spread of the second disease.}
\label{sketch}
\end{figure}

Now consider the case in which our first disease causes an epidemic in the
network, leaving a fraction~$S$ of vertices either dead or immune to
infection by our second disease (or by a second wave of the first disease).
To represent this mathematically, we remove these vertices from the
network, leaving a smaller network of uninfected vertices which we call the
\defn{residual graph}---see Fig.~\ref{sketch}.  Only if this residual graph
has a giant component will it be possible for the second pathogen, provided
it has a suitably high transmissibility, to spread.

Clearly when $T=0$ for the first pathogen no individuals are infected and
the entire graph remains for the second pathogen to exploit.  Conversely,
when $T=1$ an epidemic of the first pathogen will infect the entire giant
component of the graph, and once this component is removed the second
pathogen definitely cannot spread (since, in the limit of large size,
random networks have only one giant component).  In between these two
extremes, we can expect a transition, which we now investigate.  We begin
by calculating the degree distribution of the residual graph.  Once we have
this distribution then, because the graph is uncorrelated, it is a
straightforward exercise to determine whether it has a giant component or
not.

Consider a vertex with degree~$k$.  Let $P(\mbox{uninf.},m|k)$ be the
probability that it remains uninfected at the end of the first epidemic and
has $m$ edges that are attached to other uninfected vertices.  In other
words, $P(\mbox{uninf.},m|k)$ is the probability that this vertex belongs
to the residual graph and has degree~$m$ within that graph, given that it
has degree~$k$ in the graph as a whole.

This probability is equal to the probability that the vertex has $k-m$
unoccupied edges that attach to infected vertices and $m$ edges (occupied
or not) that attach to uninfected vertices.  The probability of an edge
attaching to a uninfected vertex is just $F_1(u)$ and the probability of
being unoccupied and attaching to an infected vertex is $(1-T)(1-F_1(u)) =
u - F_1(u)$, where we have used Eq.~\eref{defsu}.  Then
$P(\mbox{uninf.},m|k) = {k\choose m}[F_1(u)]^m [u-F_1(u)]^{k-m}$.
Multiplying by the probability $p_k$ of having degree~$k$ and summing
over~$k$ then gives the probability of being uninfected and having
degree~$m$ within the graph of uninfected vertices: $P(\mbox{uninf.},m) =
\sum_{k=m}^\infty p_k {k\choose m}[F_1(u)]^m [u-F_1(u)]^{k-m}$.  Dividing
by the prior probability $P(\mbox{uninf.})=1-S=F_0(u)$ of being uninfected,
the probability distribution of the degrees of vertices within the residual
graph is
\begin{equation}
P(m|\mbox{uninf.}) = {1\over F_0(u)} \sum_{k=m}^\infty p_k
                     {k\choose m} \bigl[F_1(u)\bigr]^m
                     \bigl[u-F_1(u)\bigr]^{k-m}.
\end{equation}
The generating function for this distribution is
\begin{eqnarray}
G_0(x) &=& {1\over F_0(u)} \sum_{m=0}^\infty x^m
           \sum_{k=m}^\infty p_k {k\choose m} \bigl[F_1(u)\bigr]^m
           \bigl[u-F_1(u)\bigr]^{k-m}
           \nonumber\\
       &=& {1\over F_0(u)} \sum_{k=0}^\infty p_k
           \sum_{m=0}^k {k\choose m} \bigl[xF_1(u)\bigr]^m
           \bigl[u-F_1(u)\bigr]^{k-m}\nonumber\\
       &=& {1\over F_0(u)} \sum_{k=0}^\infty p_k
           \bigl[u+(x-1)F_1(u)\bigr]^k\nonumber\\
       &=& {F_0\bigl(u+(x-1)F_1(u)\bigr)\over F_0(u)}.
\label{defsg0}
\end{eqnarray}

Given this generating function, we can determine whether the residual
network has a giant component using the method of Ref.~\cite{NSW01}.  We
define
\begin{equation}
G_1(x) = {G_0'(x)\over G_0'(1)}
       = {F_1\bigl(u+(x-1)F_1(u)\bigr)\over F_1(u)},
\label{defsg1}
\end{equation}
which is the generating function for the excess degree of a vertex reached
by following an edge in the residual graph, precisely analogous to
Eq.~\eref{defsf1}.  Then there is a giant component if and only if
$G_1'(1)>1$.  Thus, the point $G_1'(1)=1$ constitutes an \emph{additional
phase transition} in the system, other than the standard epidemic
transition, at which a sufficiently contagious second pathogen can cause an
epidemic after the passage of the first through the network.  We call this
the \defn{coexistence transition} and the point at which it occurs the
\defn{coexistence threshold}.  Making use of Eq.~\eref{defsg1}, we find
that the transmissibility~$T_x$ at this point is the solution of the
equation
\begin{equation}
F_1'(u) = 1,
\label{defstx}
\end{equation}
where $u$ is a function of~$T$ via Eq.~\eref{defsu}.

For instance, in the case of a Poisson degree distribution for the original
network $p_k=\e^{-z}z^k/k!$ (the standard Bernoulli random graph), we have
$F_0(x)=F_1(x)=\e^{z(x-1)}$, which means that the normal epidemic threshold
falls at $T_c=1/z$ while the coexistence threshold falls at the point
satisfying
\begin{equation}
1 = z F_1(u) = z(1-S).
\end{equation}
If we can find $S$ from Eq.~\eref{gc}, it is then a straightforward matter
to find~$T_x$.

\begin{figure}
\includegraphics[width=\columnwidth]{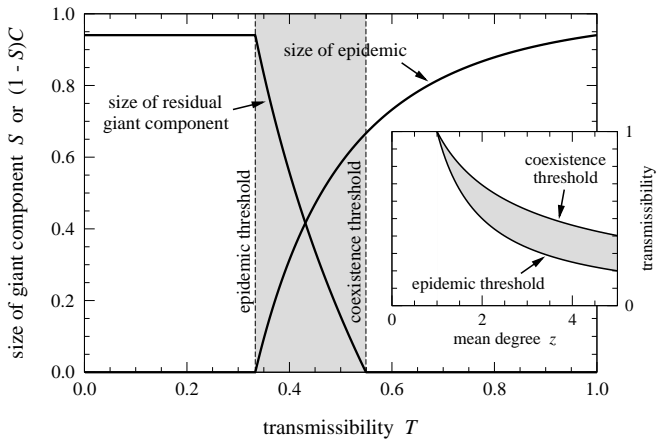}
\caption{The size of the epidemic of the first pathogen and the size of the
residual giant component that it leaves behind, as a function of
transmissibility on a graph with a Poisson degree distribution with mean
degree~$z=3$.  Inset: the position of the two thresholds as a function of
mean degree for the Poisson case.  The shaded areas in the plots denote the
region in which both pathogens can spread.}
\label{results}
\end{figure}

The size~$C$ of the giant component in the residual graph, which sets an
upper bound on the size of a possible second epidemic, is given
by
\begin{equation}
C = 1 - G_0(v),\qquad v = G_1(v),
\end{equation}
as a fraction of the size of the residual graph~\cite{NSW01}.  To get the
result as a fraction of the size of the original network, we then need to
multiply by~$1-S$.  In Fig.~\ref{results} we show the sizes $S$ and
$(1-S)C$ of the epidemic and the giant component on the residual graph as a
function of transmissibility for the Poisson case.  As the transmissibility
increases from zero, the size of the residual giant component is initially
equal to the size of the giant component of the entire graph, which is very
nearly~1.  As $T$ passes the epidemic threshold for the first pathogen,
however, the pathogen starts to spread and kills or renders immune to the
second pathogen some fraction of the population, thereby reducing the size
of the epidemic of the second pathogen.  At some point---our coexistence
threshold---so many are killed or made immune that too few are left to
spread the second pathogen and $C$ reaches zero.  Thus the epidemic spread
of both pathogens is possible only in the intermediate regime of
transmissibility $T_c<T<T_x$ indicated by the shaded area in the figure; if
the transmissibility is either too low or too high, coexistence is
impossible.  In the inset of the figure we show how the two threshold
values of the transmissibility, $T_c$~and~$T_x$, vary as a function of mean
degree for the Poisson case.

Of course, the mere existence of a giant component in the residual graph
does not mean that the second pathogen will cause an epidemic.  That
depends on whether the transmissibility of the second pathogen is high
enough.  Repeating the analysis leading to Eq.~\eref{tc}, we find that the
second pathogen can spread if its transmissibility is above the critical
value $T_c'=1/G_1'(1)$ or equivalently $T_c'=1/F_1'(u)$---yet a third
threshold in our system (but one whose position is not solely a function of
the network topology, since it depends also on the transmissibility of the
first pathogen via Eq.~\eref{defsu}).  Noting that $F_1'(x)$ is a
polynomial with non-negative coefficients and therefore monotonic
increasing on the positive real line (within its radius of convergence),
and that $u\le1$ since it is a probability, we see that $F_1'(u)\le
F_1'(1)$, and hence that $T_c'\ge T_c$: the minimum transmissibility
necessary for the second pathogen to spread is never less than that
necessary for the first.  This accords with our intuition: as we have shown
elsewhere~\cite{Newman02c}, vertices with high degree are more likely to be
infected than those with low degree, and therefore we would expect the
residual graph to have lower mean degree and hence higher epidemic
threshold than the original network.

Another example of interest is that of a network with a power-law degree
distribution $p_k=k^{-\alpha}/\zeta(\alpha)$, for some constant~$\alpha$,
where $\zeta(x)$ is the Riemann $\zeta$-function.  Such networks are often
called ``scale-free.''  A variety of networks appear to be scale-free and
they have attracted considerable attention in the recent
literature~\cite{AB02,DM02}.  As is by now well
understood~\cite{CEBH00,PV01a}, (uncorrelated) scale-free networks with
$\alpha<3$ have a vanishing epidemic threshold $T_c=0$ because the second
moment $\av{k^2}$ of the degree distribution in Eq.~\eref{threshold}
diverges.  Noting~\cite{NSW01} that a power-law degree distribution gives a
generating function $F_0(x)=\Li_\alpha(x)/\zeta(\alpha)$, where $\Li_n(x)$
is the $n$th polylogarithm of~$x$, and applying Eq.~\eref{defstx}, we find
by contrast that the \emph{coexistence} threshold in such a network is in
general nonzero.  Furthermore, the critical transmissibility for the second
pathogen $T_c'=1/F_1'(u)$ is also nonzero.  (This nonzero threshold
immediately implies that the residual network cannot itself be scale-free.
The physical explanation of this result is that the first pathogen is more
likely to infect higher-degree vertices and so selectively removes or
immunizes the ``hubs'' in the network, destroying the power-law form.
Removing hubs is well-known to be a good strategy for preventing the spread
of disease~\cite{DB02,CBH03}.)

The result $T_c=0$ implies that a single disease spreading on a scale-free
network of this kind can never be eradicated by an intervention whose sole
effect is to reduce the transmissibility.  Our findings indicate, however,
that for the case of two competing pathogens on such a network, \emph{one
of them} can be eradicated by an intervention that lowers the
transmissibility, but not both.

To conclude, we have studied, using mappings to bond percolation, the
problem of two diseases spreading through the network of contacts between
members of a host population.  We find that, in the case where hosts can be
infected with either one or other, but not both, of the diseases, the
spread of both is possible only for intermediate values of the
transmissibility of the first disease.  There are two phase transitions
that mark the boundaries of this intermediate regime.  The first is the
standard epidemic transition below which the first disease is not
contagious enough to spread at all; the second is an additional topological
phase transition in the network that corresponds to the point at which the
first disease removes from the population so large a fraction of the hosts
that not enough remain to support the spread of the second disease.

We have here studied only the simplest case of competing pathogens.  A
number of variants of the problem are of interest.  For instance, in some
cases the first pathogen may confer upon those it infects only partial
cross-immunity to the second, so that the probability of infection with the
second pathogen is reduced but not entirely eliminated.  This process could
be modeled using an extension of the formalism described here in which the
residual graph is formed by removing a fixed fraction, randomly selected,
of the vertices affected by the first epidemic.

The author thanks Ben Kerr, James Koopman, Mercedes Pascual, and Carl Simon
for useful conversations.  This work was funded in part by the National
Science Foundation under grant number DMS--0405348.

\end{document}